\documentclass[aps,twocolumn,groupedaddress,floats,showpacs]{revtex4}
\usepackage[dvips]{graphicx}
\usepackage{bm}
\begin{document}

\title{Ginzburg-Landau-Wilson Aspects of Deconfined Criticality}
\author{Anatoly Kuklov${^1}$,
Nikolay Prokof'ev$^{2,3}$, and Boris Svistunov$^{2,3}$}
\affiliation{ ${^1}$Department of Engineering Science and
Physics, The College of Staten Island, City University of New
York, Staten
Island, NY 10314\\
${^2}$Department of Physics, University of
Massachusetts, Amherst, MA 01003 \\
${^3}$Russian Research Center ``Kurchatov Institute'', 123182
Moscow }

\begin{abstract}
Monte Carlo study of the deconfined critical action  phase diagram
reveals a region where spinon deconfinement occurs through a weak
first-order phase transition, in agreement with Ginzburg-Landau
theory. Wilson renormalization argument in combination with the
absence of the data collapse even in the regime of weak
interaction between the spinons casts a serious doubt on the
possibility of the continuous deconfinement transition. We also
argue that if a continuous deconfined criticality does exist on
the phase diagram, its nature is analogous, in a certain precise
sense, to that of the XY universality class.
\end{abstract}

\maketitle

Critical properties of systems described by one or several complex
fields coupled to one gauge field have a long history of studies
and apply to numerous problems in physics which include
normal-superfluid transitions in multi-component neutral or charged
liquids (see, e.g. \cite{Babaev}), superfluid--valence-bond solid
(SF--VBS) transitions in lattice models (\cite{dcp1,dcp2,dcp3}),
Higgs mechanism in particle physics \cite{Higgs}, etc. Recently,
the authors of Refs.~\cite{dcp1,dcp2} argued that the SF--VBS
transition in a $(2+1)$-dimensional system is an example of a
qualitatively new type of quantum criticality that does not fit
the Ginzburg-Landau-Wilson (GLW) paradigm.
The discussion of this claim is the main focus of our Letter.

The theory of deconfined critical point (DCP) establishes a
remarkable microscopic picture of how a generic continuous SF--VBS
transition may happen. A generic II-order SF--VBS transition can
not be derived from na\"{\i}ve GLW expansion in powers of
order parameters starting from the quantum disordered groundstate,
which has no broken symmetries, simply because such a state is
unlikely to be present in most models \cite{weak1,dcp3}. It may
happen, however, that restrictions imposed on parameters of the
GLW action are automatic and originate from hidden (or emerging)
symmetries. In the DCP action the crucial symmetry of this kind is
that between the spinons (or vortices) in the VBS state. Another
alternative is that the DCP action {\it in fact} predicts that the
SF-VBS phase transition is of the I-order.

In what follows we discuss results of the Worm algorithm Monte
Carlo simulations performed for the DCP action, construct its
phase diagram, and show that the SF--VBS transition is I-order at
least for sufficiently strong coupling between the spinon fields.
This explains the outcome of recent numerical simulations which
report either difficulties with finite-size scaling of the data
\cite{Sandvik} or weak I-order transitions \cite{weak1}. It is far
more difficult to present a direct evidence of the I-order
transition in the weak coupling limit. However, problems with the
data collapse in finite-size systems hint at this possibility. The
global I-order scenario would not be a surprise in view of a
simple renormalization argument that in the DCP action the weak
coupling regime maps onto the intermediate-coupling one, where we
find a circumstantial evidence that the II-order scenario fails.

Many features of the DCP action are remarkably similar,
both qualitatively and quantitatively, to those of a more
conventional two-component XY-model (2XY) \cite{twocolor}.
Moreover, we argue that if the line of II-order DCP's does exist,
then its critical properties are most close to those of 2XY at the
U(1)$\times$U(1) critical point, in the following sense. There
exists a self-dual model with {\it marginal} long-range
interactions $\propto 1/r^2$, which continuously interpolates
between DCP and U(1)$\times$U(1) criticality by varying amplitudes
of $\propto 1/r^2$ terms.  

There are many equivalent formulations of the action describing
coupled gauge and multi-component complex fields  
\cite{dcp1,Babaev,Motrunich}. Here we employ the integer-current
lattice representation which can be viewed either as a
high-temperature expansion for the XY model in three
dimensions, or as a path-integral (world-line) representation of
the interacting quantum system in discrete imaginary time in
$(d+1)=3$ dimensions. The XY action reads
\begin{equation}
S = U_{{\bf r} - {\bf r}'}{\bf j}_{{\bf r}}\cdot {\bf j}_{{\bf r}'} \;, \label{XY}
\end{equation}
where summation over repeated lattice sites ${\bf r}$ is assumed;
$U_{{\bf r} - {\bf r}'}=U\delta_{{\bf r},{\bf r}'}$; and ${\bf
j}_{\bf r} =  (j_{{\bf r}})_{\mu} $ with $\mu = x, y, \tau $ are
integer, zero-divergence, $\nabla {\bf j} = 0$, currents defined
on bonds of the simple cubic space-time lattice with periodic
boundary conditions. The configuration space of $j$-currents is
that of closed oriented loops. In terms of particle world lines,
currents in the time direction represent occupation number
fluctuations away from the commensurate filling, and currents in
the spatial directions represent hopping events. The state of $S$
with small current loops is normal (``high-temperature"); the
corresponding particle order-parameter, $\psi$, is zero in this
phase. When particle world lines proliferate and grow
macroscopically large, the system enters into the superfluid state
with $\langle \psi \rangle \ne 0$.

Generalization to the symmetric multi-component case with short-range
interactions is straightforward (in what follows we will
concentrate on the two-component case which describes coupled
XY models)
\begin{equation}
S_2^{(s)} =U\: {\bf j}_{1{\bf r}}^2 +
U \: {\bf j}_{2{\bf r}}^2 -
V \: \left( {\bf j}_{1{\bf r}}+{\bf j}_{2{\bf r}} \right)^2 \;.
\label{AB}
\end{equation}
If particle/XY-spin fields in the original model are coupled
through a gauge field, then after the gauge field is integrated
out one arrives at the DCP action similar to Eq.~(\ref{AB}), which
now contains a long-range part \cite{Babaev,Motrunich}
\begin{equation}
S_2^{(l)} =U {\bf j}_{1{\bf r}}^2 + U {\bf j}_{2{\bf r}}^2 +
g Q_{{\bf r}- {\bf r}'}
 \left( {\bf j}_{1{\bf r}}+{\bf j}_{2{\bf r}} \right) \cdot
 \left( {\bf j}_{1{\bf r}'}+{\bf j}_{2{\bf r}'} \right) \;.
\label{DCPA}
\end{equation}
The lattice Fourier transform of the interaction potential
$Q_{{\bf r}- {\bf r}'}$ is given by  $Q_{\bf q} = 1/\sum_{\mu }
\sin^2 (q_{\mu}/2)$, which implies the asymptotic form $Q(r \to
\infty ) \sim 1/r$. In the discussion of the SF--VBS transition,
$j_1$ and $j_2$ currents in the DCP action represent world lines
of spinons which are VBS vortices carrying fractional particle
charge $\pm 1/2$ \cite{dcp1,dcp2}.

\begin{figure}[tbp]
\includegraphics[bb=70 205 575 585, angle=-90, width=0.55\columnwidth]{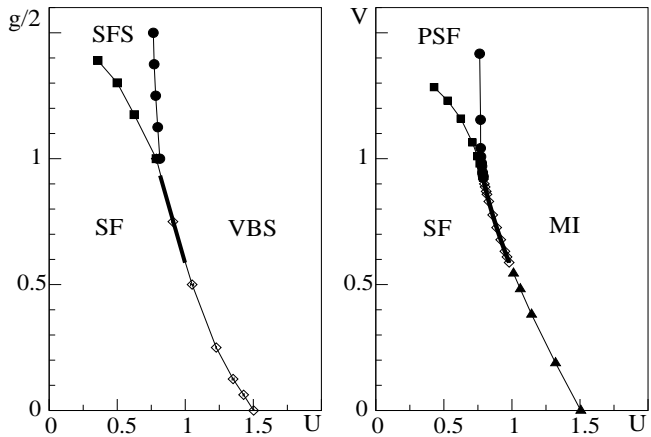}
\caption{Phase diagrams of the long-range (left panel),
Eq.~(\ref{DCPA}), and short-range (right panel), Eq.~(\ref{AB}),
actions. Bold solid lines indicate regions where I-order phase
transitions were unambiguously verified. Error bars are shown but
are typically much smaller than symbol sizes.} \label{fig1}
\end{figure}

Phase diagrams for the long- and short-range actions are shown in
Fig.~\ref{fig1}. Different phases are identified in terms of loop
sizes for $j_1$ and $j_2$ currents as follows: the Mott insulator
(MI) and VBS states are characterized by small $j_1$- and
$j_2$-loops; in the VBS supersolid (SFS) and paired superfluid
(PSF) states only $(j_1 \mp j_2)$-loops grow macroscopically large
while single component loops remain small; finally, in the SF
state there are macroscopically large $j_1$- and $j_2$-loops. The
data for the short-range model are reproduced from
Ref.~\cite{twocolor}. A similar shape of the phase diagram for the
DCP action in the angle-gauge field representation was obtained in
Ref.~\cite{motrunich2}, though, no I-order transition was
identified \cite{note2}.

\begin{figure}[tbp]
\includegraphics[bb=0 220 570 570, angle=-90, width=0.45\columnwidth]{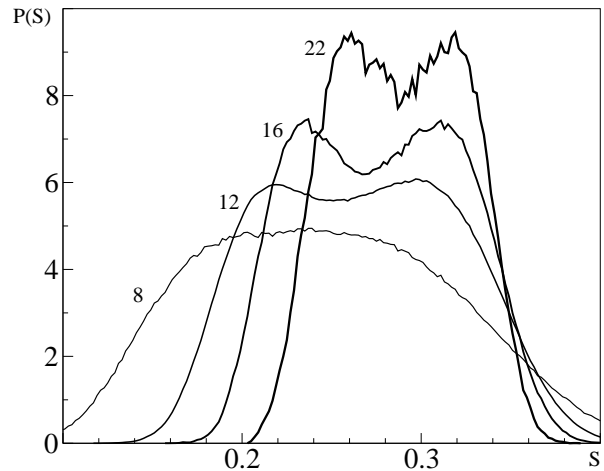}
\caption{Normalized probability density distributions $P(S)$ of the DCP action
for $g=1.5$ and different system sizes; U(L=8)=0.94375, U(L=12)=0.92345,
U(L=16)=0.91667, U(L=22)=0.91274. }
\label{fig2}
\end{figure}

It is hard not to notice a remarkable {\it quantitative}
similarity between the phase diagrams. In particular, the DCP
action has VBS to SFS and SFS-SF transitions when coupling between
the spinons is strong and they form tightly bound particle states.
One immediate prediction based on properties of the short-range
model would be that close to the SF-SFS-VBS bicritical point the
SF-VBS transition is I-order in nature. Unfortunately, the I-order
transition in $S_2^{(s)}$ is very weak and can be clearly seen
only at system sizes $L>24$ by calculating the probability density
for the system action, $P(S)$, and observing its double-peak
structure. Simulating large $L$ for the long-range action is far
more difficult. We were able to collect reliable statistics for
$S_2^{(l)}$ only for $L\le 22$. Still, for $g=1.5$ we resolved
(Fig.~\ref{fig2}) the development of the double-peak structure in
$P(S)$ starting from an anomalously flat maximum at $L=8$.

\begin{figure}[tbp]
\includegraphics[bb=0 190 570 590, angle=-90, width=0.45\columnwidth]{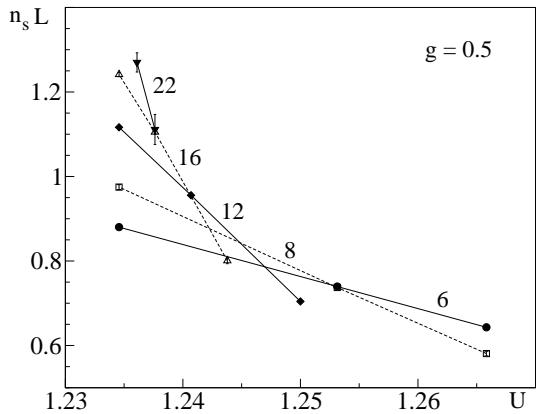}
\caption{ $n_s (U) L$ curves for $g=0.5$ and system sizes $L=6,8,12,16,22$.}
\label{fig3}
\end{figure}

For small $g$ one expects the I-order transition to become weaker
(the $g=0$ point describes two independent XY models) and nearly
impossible to study using $P(S)$ functions. However, there are two
important circumstances. (i) At $g \to 0$, the initial part of the
renormalization flow with $L \to \infty$ simply leads to
increasing $g(L) \propto L$ \cite{Halperin2} [see also
Eq.~(\ref{RG})], thus mapping the weak-coupling regime onto the
intermediate-coupling one. [Indeed, $1/r$ interaction is a
relevant operator responsible for the spinon confinement.] (ii)
For the superfluid stiffness, $n_s$, a scale invariant II-order
criticality implies $n_s L  \to {\rm const}$ as $L \to \infty$.
These facts has to be combined with the numerical observation  
that for $g>0.1$ the $ n_s L $ curves do not intersect at the 
same $U$ even approximately (see Fig.~\ref{fig3}). This strongly 
suggests the I-order transition scenario for all $g$ (we exclude 
a {\it non-scale-invariant} II-order scenario).

{\it Critical self-duality of XY models.~} Speaking strictly, one
cannot rule out that a II-order line actually starts below $g =
0.1$, or, that for some reason the data collapse for $g=0.5$ sets
in only at $L \gg 20$. A possibility of II-order phase transitions
in generic multi-component XY models with $1/r$ current-current
interactions can hardly be questioned since the duality
transformation \cite{Halperin} for the short-ranged action
$S_2^{(s)}$ produces a family of such models. To be unambiguous in
our further analysis, we adopt the following two definitions. \\
{\it Def.~1}: We call a model self-dual if there exists a duality
transformation which at the critical point preserves the form of
the Hamiltonian, but not necessarily the values of the critical
parameters. \\
{\it Def.~2}: A model ${\cal A}$ is critically self-dual if
there exists a self-dual model ${\cal B}$ such that the critical
properties of ${\cal A}$ and ${\cal B}$ are the same.

XY action can be identically transformed into the dual
(inverted-XY) action \cite {Halperin} which has the same
functional form as Eq.~(\ref{XY}) for integer, zero-divergence
currents ${\bf l}$ coupled by the long-range dual interaction
potential, $U^{(d)}_{{\bf r} \to \infty} \sim 1/r$. In Fourier
space, the dual potential is given by $ U^{(d)}_{\bf q}=(\pi^2 /
4) Q_{\bf q}/U$. The notion of bond currents also changes from
particle to vortex world lines. Now, the proliferation of vortex
${\bf l}$-lines signals the onset of the transition to the normal
state. Formally, long-range interactions in the dual action make
the XY model not self-dual in conventional sense, though the
configuration space is preserved. Several papers suggested that
$1/r$ interactions result in the inverted-XY criticality
qualitatively different from the conventional XY point
\cite{Herbut,hove}.

A single-component short-range XY model is critically self-dual.
Below we prove this statement numerically. Meanwhile, it is easy
to show analytically that XY criticality is {\it arbitrarily
close} to the critically of self-dual models. First, we introduce XY
models with $1/r^2$-potential, which are self-dual. Indeed, the
potential with the Fourier transform $U_{{\bf q}\to 0} \to A (\pi
/2) \sqrt{Q(q)} + B + \dots $ under duality transformation becomes
$U^{(d)}_{{\bf q}\to 0} \to A^{-1} (\pi /2) \sqrt{Q(q)} -B/A^2 +
\dots$. Note that the amplitude in front of the $1/r^2$ term
changes.  Second, we observe that the $1/r^2$ interaction is {\it
marginal} (this will be explicitly seen below) and thus by adding
a $1/r^2$-term with infinitely small amplitude to XY model
(\ref{XY}) we do not change its critical behavior and exponents,
but formally make it self-dual. It is of crucial importance that
XY criticality is mapped onto that of a model with {\it marginal}
interaction; this is required to reconcile critical self-duality
with different critical exponents and universal numbers of the XY
and inverted XY transitions. These differences naturally follow
from different amplitudes of the dual $1/r^2$-terms. The XY and
inverted XY critical points turn out to be just two points in 
the {\it continuum} of $A/r^2$-criticalities, corresponding to 
two special values, $A_-$ and $A_+$. By changing $A$ one may 
continuously interpolate between $A_-$ and $A_+$, gradually 
transforming XY criticality into inverted XY, and vice versa.

To illustrate the above-mentioned point and, in particular, to
extend the proof to the case of finite $A$, we computed the
critical exponent $D_H(A)$ for the interaction potential $U_{{\bf
q}} = A (\pi /2) \sqrt{Q(q)} + B$. The Hausdorff dimension $D_H$
gives the the average line length of critical loops as a function
of system size, $\langle l \rangle \propto L^{D_H}$. Under duality
transformation. $D_H(A)$ transforms into $D^{(d)}_H(A)$. As
expected, among a continuum of self-dual critical points $B_c(A)$
there is one, $A\approx 3.2$, $B_c(A=3.2)=-2.8457(10)$ which
reproduces Hausdorff dimensions of the inverted-XY and XY systems
respectively within the statistical error bars  \cite{note1}.

To understand the origin of critical self-duality, and in
particular, the physical meaning of the $1/r^2$ interaction, we
start with constructing proper coarse-grained current variables
similar to winding numbers in finite-size systems and subject to
the real space-time renormalization group (RG) transformations.
Crucial for our derivation will be the fact that at criticality
there are $\sim {\cal O}(1)$ loops of size $\sim r$ in the volume
$r^3$ \cite{Williams}, which is nothing but the requirement of
scale invariance.

Imagine a cube of linear size $u$ centered at point ${\bf r}$,
where $u \gg 1$ is odd integer. Its facets cut bonds of the
original lattice a distance $u/2$ away from the center. Integer,
zero-divergence currents ${\bf J}_{\bf r}$ are defined as current
fluxes through the facets of the cube (we denote the cube by
$C_{\bf r}$ and its facets as $S_{{\bf r}, \mu}$)
\begin{equation}
(J_{\bf r})_{\mu}=\sum_{{\bf r}' \in S_{{\bf r}, \mu}} (j_{{\bf
r}'})_{\mu}\;, \label{facet}
\end{equation}
with an additional restriction that only $j$-loops of size larger
than $u/2$ contribute to the sum (\ref{facet}). The last condition
is necessary to suppress noise due to small loops winding around
the cube edges. The new variables are scale-invariant by
construction, i.e. $\langle J^2 \rangle \sim {\cal O}(1)$. This
fact alone allows us to study the RG flow of the long-range
effective interaction potential. Let the original model be
long-range with the bare potential decaying as $g/r^{\alpha},\,
\alpha \leq  2$. After coarse-graining the interaction energy
between two cubes $C_{{\bf r}_1}$ and $C_{{\bf r}_2}$ separated by
a distance $R=|{\bf r}_1- {\bf r}_2| \gg u$ can be estimated as
\begin{equation}
{g \over \epsilon (u)R^{\alpha }} \: 
\sum_{{\bf x} \in C_{{\bf r}_1} } \sum_{{\bf y} \in C_{{\bf r}_2} }
{\bf j}_{{\bf x}} \cdot {\bf j}_{{\bf y}} \sim
{g \: u^2 \over \epsilon (u)R^{\alpha } } {\bf J}_{{\bf r}_1}\cdot {\bf J}_{{\bf r}_2} \;,
\label{RG}
\end{equation}
where $\epsilon (u)$ is the ``dielectric" constant at length-scale $u$.
The last relation follows from the continuity of world lines and
the fact that ${\bf J}$-currents are facet, not bulk, sums, so
that $\sum_{{\bf x} \in C_{{\bf r}_1} }{\bf j}_{{\bf x}}\sim u
{\bf J}_{{\bf r}_1}$. We now rescale all distances by a factor of
$u$ and observe that the effective interaction potential
transforms as $ g u^{2-\alpha}/\epsilon (u) r^{\alpha }$. 
This means that (i) long-range interactions 
with $\alpha <2$ are screened at the fixed point, 
$\epsilon (u) \propto u^{2-\alpha} \to 0$, and (ii) the term
${\bf J}_{{\bf r}}\cdot{\bf J}_{{\bf r}'}/({\bf r}- {\bf r}')^2$
is a marginal operator. 
For $\alpha =1$ and small initial values of $g$, the RG flow starts 
with $epsilon =1$ and leads to $g(u)\sim u$. This result was used 
to relate the weak- and intermidiate-coupling regimes in the DCP action. 
This consideration does not depend on the number of components, but
it works only as long as $g(u)$ remains small. 

The above analysis is supported by microscopic mechanisms 
leading to the renormalized $1/r^2$-law between microscopic currents. 
In the SF phase, the $1/r$ potential originates from the
kinetic energy of the superflow around vortices. Close to the
critical point, the superflow around a vortex line is screened on
large distances by small vortex loops, and the kinetic energy
integral is transformed into (for simplicity we consider here a
straight vortex line) $\int n_s(r) dr /r \to \int dr /r^2$ with
scale-dependent superfluid density $n_s\propto 1/r$. This formula 
implies that at the critical point the interaction energy 
between the vortex line elements is proportional to $1/r^2$. 
The emergence of the renormalized potential ~$\sim 1/r^2$ 
while approaching the critical point from the normal phase, 
can be considered as a result of virtual exchange of long-wave
sound excitations. Elementary calculations in the hydrodynamical
approximation for $d=2$ superfluid show that such potential scales
as $\sim 1/r^2$.

One may also take an alternative, mathematically rigorous, point of
view and consider an action describing a {\it single} $j$-current line 
(of arbitrary fractal structure) at the critical point when other 
degrees of freedom in the system are integrated out
\begin{equation}
S_{\rm crit}^{(\rm {loop})} = \bigg[ (U_{\rm crit})_{{\bf r}- {\bf
r}'} +\lambda \: \delta_{{\bf r}- {\bf r}'} \bigg] \: {\bf
j}_{{\bf r}} \cdot{\bf j}_{{\bf r}'} \; , \label{polymer}
\end{equation}
where $U_{\rm crit} = A/r^2 + \mbox{short-range terms} $.
It is written in terms of original $j$- or $l$-currents and 
{\it not} coarse-grained variables. In particular, the statistics 
of the disconnected loop, formally identical to a polymer of a 
variable length, must reproduce geometrical exponents at the critical 
point. It is this critical-loop action that unifies short- and 
long-range XY models, putting them in a wider context of self-dual 
$1/r^2$ models.

In conclusion, we present numerical and RG agruments
strongly suggesting that the DCP action is the theory of 
weak I-order SF-S transitions.
In the ``zoo" of self-dual $1/r^2$ criticalities interpolating between
short-range and long-range $1/r$ models, the number of
components does not seem to play any qualitative role unless the
transition turns into the I-order one. 

We are grateful to E. Vicari, O. Motrunich, S. Sachdev, M. Fisher,
and Z. Te\v{s}anovi\'{c} for valuable and stimulating discussions.
The research was supported by the National Science Foundation
under Grants No. PHY-0426881 and PHY-0426814, and by PSC CUNY
Grant No. 665560035.

\end{document}